\documentclass[aps,twocolumn,showpacs,preprintnumbers,floats]{revtex4}\def\@cite#1#2{\textsuperscript{[{#1\if@tempswa , #2\fi}]}}
\usepackage{mathrsfs}
\usepackage{longtable,lscape}
\usepackage{txfonts}
\usepackage{amssymb}
\usepackage{indentfirst}
\usepackage{graphicx,booktabs}
\usepackage{float}
\usepackage{setspace}

\usepackage{multirow}
\usepackage{color}
\usepackage{epsfig}
\newcommand{\vsig}{\mbox{\boldmath$\sigma$\unboldmath}}

\begin{document}


\title{A possible interpretation of the newly observed $\Omega(2012)$ state}

\author{Li-Ye Xiao$^{1,2}$~\footnote {E-mail: lyxiao@pku.edu.cn} and Xian-Hui Zhong$^{3,4}$~\footnote {E-mail: zhongxh@hunnu.edu.cn}}

\affiliation{ 1) School of Physics and State Key Laboratory of
Nuclear Physics and Technology, Peking University, Beijing 100871,
China } \affiliation{ 2)  Center of High Energy Physics, Peking
University, Beijing 100871, China} \affiliation{ 3) Department of
Physics, Hunan Normal University, and Key Laboratory of
Low-Dimensional Quantum Structures and Quantum Control of Ministry
of Education, Changsha 410081, China } \affiliation{ 4) Synergetic
Innovation Center for Quantum Effects and Applications (SICQEA),
Hunan Normal University, Changsha 410081, China}


\begin{abstract}

Inspired by the newly observed $\Omega(2012)$ state at Belle II, we investigate the two-body
strong decays of $\Omega$ baryons up to $N=2$ shell within the chiral quark model.
Our results indicate that: (i) the newly observed $\Omega(2012)$ state could be
assigned to the spin-parity $J^P=3/2^-$ state $|70,^210,1,1,\frac{3}{2}^-\rangle$
and the experimental data can be reasonably described. However, the spin-parity $J^P=1/2^-$
state $|70,^210,1,1,\frac{1}{2}^-\rangle$ and spin-parity $J^P=3/2^+$ state $|56,^410,2,0,\frac{3}{2}^+\rangle$
can't be completely excluded. (ii) The $D$-wave states in the $N=2$ shell are most likely to be narrow states
with a width of dozens of MeV and have a good potential to be observed in the $\Xi K$ and/or $\Xi(1530)K$
channels in future experiments. The $\Omega(2250)$ resonance listed in PDG may be a good
candidate of the $J^P=5/2^+$ $1D$ wave state $|56,^410,2,2,5/2^+\rangle$.

\end{abstract}

\pacs{}

\maketitle

\section{Introduction}

Searching for the missing baryon resonances and understanding the baryon spectrum are important topics in hadron physics.
In the past years, for the limitations of experimental conditions, our knowledge about $\Omega$ spectrum is still scarce.
There are only a few data on the $\Omega$ resonances. In the review of the Particle Data Group (PDG)~\cite{Patrignani:2016xqp},
only four $\Omega$ baryon states are listed: $\Omega(1672)$, $\Omega(2250)$, $\Omega(2380)$, and $\Omega(2470)$.
Except the ground state $\Omega(1672)$ being well established with four-star ratings,
the nature of the other three excited states are still rather uncertain with three- or two-star ratings.
Fortunately, the Belle II experiments offer a great opportunity for our study of the $\Omega$ spectrum.

Very recently, Belle II Collaboration reported a new excited hyperon, an $\Omega^{*-}$ candidate
(denoted by $\Omega(2012)$ here), in the $\Xi^0K^-$ and $\Xi^-K^0$  mass distributions very recently~\cite{Yelton:2018mag}.
Its mass and decay width are measured to be
\begin{eqnarray}
M=2012.4\pm0.7~(\text{stat})\pm0.6~(\text{syst})~\text{MeV}
\end{eqnarray}
and
\begin{eqnarray}
\Gamma=6.4^{+2.5}_{-2.0}~(\text{stat})\pm0.6~(\text{syst})~\text{MeV},
\end{eqnarray}
respectively.
In various models and methods, such as the Skyrme model~\cite{Oh:2007cr}, quark model~\cite{Capstick:1986bm,Faustov:2015eba,Loring:2001ky,Liu:2007yi,Chao:1980em,Chen:2009de,An:2013zoa,Kalman:1982ut,Pervin:2007wa,An:2014lga},
lattice gauge theory~\cite{Engel:2013ig,Liang:2015bxr} and so on~\cite{Carlson:2000zr,Goity:2003ab,Schat:2001xr,Matagne:2006zf,Bijker:2000gq},
the masses of the first orbital excitations of $\Omega(1672)$ are predicted to be $\sim 2.0$ GeV.
Thus, the newly observed $\Omega(2012)$ may be a good candidate of the first orbital ($1P$) excitations
of $\Omega(1672)$. In Ref.~\cite{Chao:1980em}, the mass of $2S$ state (the first radially
excitation of $\Omega(1672)$) with $J^P=3/2^+$ was predicted to be 2065 MeV, which is
also close to the mass of $\Omega(2012)$. To further determine the spin-parity quantum numbers
and inner structure of $\Omega(2012)$, besides its mass one should study its other properties,
such as magnetic moments, radiative and strong decays properties, as well.
In the early literature, limited discussions exist on the magnetic moments~\cite{Narodetskii:2013cxa},
radiative decays~\cite{Kaxiras:1985zv} of the $\Omega$ resonances.
In present work we will attempt to understand the inner structure of this newly observed state
$\Omega(2012)$ by analyzing the strong decay properties of its possible candidates
within a chiral quark model, so that our theoretical widths can be compared with the
measured width directly.

\begin{table*}[ht]
\caption{The theoretical masses (MeV) and spin-flavor-space wavefunctions of baryons, denoted by
$|\mathbf{N}_{6},^{2S+1}\mathbf{N}_{3},N, L,J^P\rangle$~\cite{Xiao:2013xi}. The
Clebsch-Gordan series for the spin and angular-momentum addition
$|J,J_z\rangle$= $\sum_{L_z+S_z=J_z} \langle LL_z,SS_z|JJ_z \rangle
\Psi^{\sigma }_{NLL_z} \chi_{S_z}$ has been omitted. } \label{wfL}
\begin{tabular}{cccccccccccccccc }\hline\hline
 ~~~~~~~&   ~~~~~~~&~~~~~~~&\multicolumn{4}{c}{\text{Theory}}  \\ \cline{4-7}
State ~~~~~~~& Wave function ~~~~~~~&\text{Experiment}~\cite{Patrignani:2016xqp}~~~~~~~&\cite{Faustov:2015eba}~~~~~~~&\cite{Capstick:1986bm}
~~~~~~~&\cite{Oh:2007cr}~~~~~~~&\cite{Chao:1980em}  \\
\hline
$|56,^{4}10,0,0,\frac{3}{2}^+ \rangle$~~~~~~~&$ |56,^410\rangle\Psi^s_{000} $ ~~~~~~~&1672.45~~~~~~~&1678      ~~~~~~~&1635~~~~~~~&1694 ~~~~~~~&1675              \\
\hline
$|70,^{2}{10},1,1,\frac{1}{2}^- \rangle$~~~~~~~&
$|70,^{2}{10}\rangle^\rho\Psi^\rho_{11L_z}+|70,^{2}{10}\rangle^\lambda\Psi^\lambda_{11L_z}$~~~~~~~&$\cdot\cdot\cdot$
~~~~~~~&1941 ~~~~~~~&1950~~~~~~~&1837 ~~~~~~~&2020    \\
\hline
$|70,^{2}{10},1,1,\frac{3}{2}^- \rangle$~~~~~~~& $|70,^{2}{10}\rangle^\rho\Psi^\rho_{11L_z}+|70,^{2}{10}\rangle^\lambda\Psi^\lambda_{11L_z}$  ~~~~~~~&$\cdot\cdot\cdot$~~~~~~~&2038  ~~~~~~~&2000~~~~~~~&1978 ~~~~~~~&2020     \\
\hline
$|70,^{2}{10},2,0,\frac{1}{2}^+ \rangle$~~~~~~~&$  |70,^{2}{10}\rangle^\rho\Psi^\rho_{200}+|70,^{2}{10}\rangle^\lambda\Psi^\lambda_{200}$ ~~~~~~~&$\cdot\cdot\cdot$~~~~~~~&\multirow{2}{*}{2301}   ~~~~~~~&2220~~~~~~~&\multirow{2}{*}{2140}~~~~~~~&2190  \\
$|56,^{4}{10},2,2,\frac{1}{2}^+ \rangle$~~~~~~~&$ |56,^410\rangle\Psi^s_{22L_z} $ ~~~~~~~&$\cdot\cdot\cdot$~~~~~~~&  ~~~~~~~&2255~~~~~~~&~~~~~~~&2210    \\
\hline
$|56,^{4}{10},2,0,\frac{3}{2}^+ \rangle$~~~~~~~&  $  |56,^410\rangle\Psi^s_{200} $   ~~~~~~~&$\cdot\cdot\cdot$~~~~~~~&\multirow{3}{*}{2173/2304}     ~~~~~~~&2165~~~~~~~&\multirow{3}{*}{2282}~~~~~~~&2065    \\
$|56,^{4}{10},2,2,\frac{3}{2}^+ \rangle$~~~~~~~&   $ |56,^410\rangle\Psi^s_{22L_z} $~~~~~~~&$\cdot\cdot\cdot$~~~~~~~&~~~~~~~&2280~~~~~~~&~~~~~~~&2215 \\
$|70,^{2}{10},2,2,\frac{3}{2}^+ \rangle$~~~~~~~&  $ |70,^{2}{10}\rangle^\rho\Psi^\rho_{22L_z}+|70,^{2}{10}\rangle^\lambda\Psi^\lambda_{22L_z}$  ~~~~~~~&$\cdot\cdot\cdot$~~~~~~~&  ~~~~~~~&2345~~~~~~~&~~~~~~~&2265  \\
\hline
$|56,^{4}{10},2,2,\frac{5}{2}^+ \rangle$~~~~~~~& $ |56,^410\rangle\Psi^s_{22L_z} $~~~~~~~&$\cdot\cdot\cdot$~~~~~~~&\multirow{2}{*}{2401} ~~~~~~~&2280~~~~~~~&$\cdot\cdot\cdot$~~~~~~~&2225  \\
$|70,^{2}{10},2,2,\frac{5}{2}^+ \rangle$~~~~~~~ & $ |70,^{2}{10}\rangle^\rho\Psi^\rho_{22L_z}+|70,^{2}{10}\rangle^\lambda\Psi^\lambda_{22L_z}$ ~~~~~~~&$\cdot\cdot\cdot$~~~~~~~& ~~~~~~~&2345~~~~~~~&$\cdot\cdot\cdot$~~~~~~~&2265    \\
\hline
$|56,^{4}{10},2,2,\frac{7}{2}^+ \rangle$~~~~~~~& $ |56,^410\rangle\Psi^s_{22L_z} $ ~~~~~~~&$\cdot\cdot\cdot$~~~~~~~&2332 ~~~~~~~&2295~~~~~~~&$\cdot\cdot\cdot$~~~~~~~&2210 \\
\hline\hline
\end{tabular}
\end{table*}

The chiral quark model~\cite{Manohar:1983md} is developed and
successfully used to study the Okubo-Zweig-Iizuka (OZI) allowed two-body strong
decays of the heavy-light mesons~\cite{Zhong:2008kd,Zhong:2010vq,Zhong:2009sk,Xiao:2014ura}
and baryons~\cite{Zhong:2007gp,Liu:2012sj,Xiao:2013xi,Yao:2018jmc,Xiao:2017udy,Wang:2017kfr,Wang:2017hej,Nagahiro:2016nsx}.
In this framework, the spatial wave functions of heavy baryons
are described by harmonic oscillators, and an effective chiral Lagrangian is
then introduced to account for the quark-meson coupling at the baryon-meson interaction vertex.
The light pseudoscalar mesons, i.e., $\pi$, $K$, and $\eta$,
are treated as Goldstone bosons. Since the quark-meson coupling is invariant under the
chiral transformation, some of the low-energy properties
of QCD are retained~\cite{Manohar:1983md,Li:1994cy,Li:1997gd,Zhao:2002id}.
Within the chiral quark model, the OZI allowed two-body strong decays
of $\Omega$ baryons up to $N=2$ shell are analyzed in present work.
The quark model classification for the $\Omega$ baryons and their theoretical
masses are listed in Table ~\ref{wfL}. According to our calculations, we obtain that (i) the newly observed $\Omega(2012)$ resonance could be assigned to the spin-parity $J^P=3/2^-$ state $|70,^210,1,1,\frac{3}{2}^-\rangle$ and the experimental data can be reasonably described. However, the spin-parity $J^P=1/2^-$ state $|70,^210,1,1,\frac{1}{2}^-\rangle$ and spin-parity $J^P=3/2^+$ state $|56,^410,2,0,\frac{3}{2}^+\rangle$ can't be completely excluded.(ii) The $D$-wave states in the $N=2$ shell are most likely to be narrow states with a with of dozens of MeV and have a good potential to be observed in their corresponding dominant decay channels.
The $\Omega(2250)$ resonance listed in PDG may be a good
candidate of the $J^P=5/2^+$ $1D$ wave state $|56,^410,2,2,5/2^+\rangle$.

This paper is organized as follows. In Sec. II we give a brief
introduction of the chiral quark model. we present our numerical
results and discussions in Sec. III and summarize our results in
Sec. IV.

\section{The chiral quark model}

In the chiral quark model, the effective low energy quark-pseudoscalar-meson coupling in the SU(3) flavor basis at tree level is given by~\cite{Manohar:1983md}
\begin{eqnarray}\label{coup}
H_m=\sum_j
\frac{1}{f_m}\bar{\psi}_j\gamma^{j}_{\mu}\gamma^{j}_{5}\psi_j\vec{\tau}\cdot
\partial^{\mu}\vec{\phi}_m,
\end{eqnarray}
where $f_m$ stands for the pseudoscalar meson decay constant. $\psi_j$ corresponds to the $j$th quark field in a baryon and $\phi_m$ denotes the pseudoscalar meson octet
\begin{eqnarray}
\phi_m=\pmatrix{
 \frac{1}{\sqrt{2}}\pi^0+\frac{1}{\sqrt{6}}\eta & \pi^+ & K^+ \cr
 \pi^- & -\frac{1}{\sqrt{2}}\pi^0+\frac{1}{\sqrt{6}}\eta & K^0 \cr
 K^- & \bar{K}^0 & -\sqrt{\frac{2}{3}}\eta}.
\end{eqnarray}

To match the nonrelativistic harmonic oscillator spatial wave function $\Psi_{NLL_z}$ in the calculations, we adopt a nonrelativistic form of Eq.~(\ref{coup}) and get~\cite{Zhao:2002id,Li:1994cy,Li:1997gd}
\begin{eqnarray}\label{non-relativistic-expans}
H^{nr}_{m}&=&\sum_j\Big\{\frac{\omega_m}{E_f+M_f}\vsig_j\cdot
\textbf{P}_f+ \frac{\omega_m}{E_i+M_i}\vsig_j \cdot
\textbf{P}_i \nonumber\\
&&-\vsig_j \cdot \textbf{q} +\frac{\omega_m}{2\mu_q}\vsig_j\cdot
\textbf{p}'_j\Big\}I_j \varphi_m,
\end{eqnarray}
where $(E_i,~\mathbf{p}_i)$, $(E_f,~\mathbf{p}_f)$ and $(\omega_m,~\mathbf{q})$ stand for the energy and three-vector momentum of the initial baryon, final baryon and meson, respectively. $\mathbf{\sigma}_j$ is the Pauli spin vector on the $j$th quark, and $\mu_q$ is a reduced mass expressed as $1/\mu_q=1/m_j+1/m'_j$. $\mathbf{p}'_j=\mathbf{p}_j-(m_j/M)\mathbf{P}_{\text{c.m.}}$ is the internal momentum of the $j$th quark in the baryon rest frame. $\varphi_m=e^{-i\mathbf{q}\cdot \mathbf{r}_j}$ and $e^{i\mathbf{q}\cdot \mathbf{r}_j}$ for emitting and absorbing a meson, respectively. The isospin operator $I_j$ associated with the pseudoscalar meson is given by
\begin{eqnarray}
I_j=\cases{ a^{\dagger}_j(u)a_j(s) & for $K^-$, \cr
a^{\dagger}_j(d)a_j(s) & for
$\bar{K}^0$, \cr
\frac{1}{\sqrt{2}}[a^{\dagger}_j(u)a_j(u)+a^{\dagger}_j(d)a_j(d)]\cos\theta
\cr - a^{\dagger}_j(s)a_j(s)\sin\theta & for $\eta$.}
\end{eqnarray}
Here, $a_j^{\dagger}(u,d,s)$ and $a_j(u,d,s)$ are the creation and annihilation operator for the $u,~d,~s$ quarks on $j$th quark. $\theta$ is the mixing angle of the $\eta$ meson in the flavor basis~\cite{Patrignani:2016xqp}.

For the decay processes, we select the initial-baryon-rest system in the calculations. Then, $\mathbf{p}_i=0$ and $\mathbf{p}_f=-\mathbf{q}$. The Eq.~\ref{non-relativistic-expans} can be further simplified and the partial decay amplitudes for $\mathcal{B}\rightarrow \mathcal{B}'\mathbb{M}$ can be calculated by
\begin{eqnarray}
\mathcal{M}[\mathcal{B}\to   \mathcal{B}' \mathbb{M}] =
3\left\langle \mathcal{B}'\left|\left\{G\vsig_3\cdot \textbf{q}
+h\vsig_3\cdot \textbf{p}'_3\right\}I_3 e^{-i\textbf{q}\cdot
\textbf{r}_3}\right|\mathcal{B}\right\rangle,
\end{eqnarray}
with
\begin{eqnarray}\label{ccpk}
h\equiv
\frac{\omega_m}{2\mu_q},~~~~~~~~~~G\equiv -(\frac{\omega_m}{E_f+M_f}+1),
\end{eqnarray}
where $\mathcal{B}'$ and $\mathcal{B}$ stand for the final and
initial baryon wave functions listed in Table.~\ref{wfL}.

With the derived decay amplitudes, the partial decay width for the emission of a light pseudoscalar meson is calculated by
\begin{equation}\label{dww}
\Gamma=\left(\frac{\delta}{f_m}\right)^2\frac{(E_f +M_f)|q|}{4\pi
M_i}\frac{1}{2J_i+1}\sum_{J_{iz}J_{fz}}|\mathcal{M}_{J_{iz},J_{fz}}|^2,
\end{equation}
where $J_{iz}$ and $J_{fz}$ represent the third components of the total angular momenta of the initial and final baryons, respectively. $\delta$ is a global parameter accounting for the strength of the quark-meson couplings.

In this work, the standard quark model parameters are adopted. Namely, we set $m_s=450$ MeV for the constituent quark mass. The decay constants for $K$ and $\eta$ are taken as $f_{K}=f_{\eta}=160$ MeV. The harmonic oscillator parameter $\alpha$ in the wave function $\Psi_{NLL_z}$ is adopted as $\alpha=400$ MeV. The masses of the final mesons and baryons are taken from the PDG~~\cite{Patrignani:2016xqp}. For the global parameter $\delta$, we fix its value the same as our previous study of the strong decays of $\Xi$ baryons~\cite{Xiao:2013xi}, i.e., $\delta=0.576$.

\section{Results and analysis}

Inspired by the newly observed $\Omega^{*-}$ candidate by Belle II Collaboration~\cite{Yelton:2018mag}, we carry out a systematic study of the strong decays of $\Omega$ baryons up to $N=2$ shell with a chiral quark model. Since the predicted mass of the $\Omega^-$ in the relativistic quark model~\cite{Chao:1980em} well agrees with the experimental measurement in PDG~\cite{Patrignani:2016xqp}, we adopt the predicted masses of the $\Omega$ resonances from Ref.~\cite{Chao:1980em} (see Table~\ref{wfL}) in our calculation.

\begin{figure}[]
\centering \epsfxsize=6.0cm \epsfbox{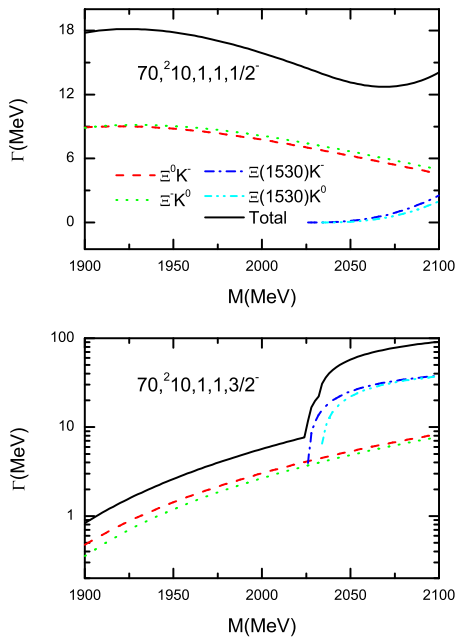} \caption{The
strong decay properties of the $1P$-wave states in the $N=1$ shell. }\label{fig-pwave}
\end{figure}

\subsection{$1P$ wave states in the $N=1$ shell }

There are two $1P$ wave states $|70,^210,1,1,1/2^-\rangle$ and $|70,^210,1,1,3/2^-\rangle$ according to the quark model classification (see Table~\ref{wfL}).  Their spin-parity quantum numbers are $J^P=1/2^-$ and $J^P=3/2^-$, respectively.
It is seen that the predicted masses in various quark models for these two $1P$ wave states are about 2000 MeV,
which are close to the mass of the newly observed $\Omega^{*-}$ candidate~\cite{Yelton:2018mag}.
As the possible assignments of the newly observed $\Omega(2012)$ state, it is crucial to study the decay properties of the two states.

Considering the $\Omega(2012)$ state as candidates of both the $J^P=1/2^-$ and $J^P=3/2^-$ states,
we calculate their strong decay properties, our results are shown in Table~\ref{pwave-num}.
As a candidate of the $J^P=3/2^-$ state $|70,^210,1,1,3/2^-\rangle$,
the predicted width
\begin{eqnarray}
\Gamma_{\mathrm{total}}^{\mathrm{th}}[\Omega(2012)]=6.6 ~\text{MeV},
\end{eqnarray}
and branching fraction ratio
\begin{eqnarray}
\mathcal{R}=\frac{\mathcal{B}[\Omega(2012)\to \Xi^0K^-]}{\mathcal{B}[\Omega(2012)\to \Xi^-\bar{K}^0]}\simeq 1.12,
\end{eqnarray}
are highly consistent with the measured width $\Gamma^{\text{exp}}=6.4^{+2.5}_{-2.0}\pm0.6$ MeV and ratio
$\mathcal{R}^{\text{exp}}=1.2\pm 0.3$ of the newly observed $\Omega(2012)$ state, .

Meanwhile, as candidate of the $J^P=1/2^-$ state $|70,^210,1,1,1/2^-\rangle$,
the predicted width for $\Omega(2012)$ is
\begin{eqnarray}
\Gamma_{\mathrm{total}}^{\mathrm{th}}[\Omega(2012)]=15.2 ~\text{MeV},
\end{eqnarray}
and the branching fraction ratio is predicted to be
\begin{eqnarray}
\mathcal{R}=\frac{\mathcal{B}[\Omega(2012)\to \Xi^0K^-]}{\mathcal{B}[\Omega(2012)\to \Xi^-\bar{K}^0]}\simeq 0.95,
\end{eqnarray}
The predicted total decay width of is about 2.5 times larger than the measured width of $\Omega(2012)$.
Considering the model uncertainties, the possibility as the assignment of  $J^P=1/2^-$ state
$|70,^210,1,1,1/2^-\rangle$ can't be excluded.

\begin{table}[htbp]
\caption{The predicted partial and total decay widths (MeV) of the assignments $|70,^210,1,1,1/2^-\rangle$ and $|70,^210,1,1,3/2^-\rangle$ with a mass of $M=2012$ MeV.} \label{pwave-num}
\begin{tabular}{ccccccc }\hline
\hline
States                   ~~~~~~&$\Gamma[\Xi^0K^-]$~~~~~~&$\Gamma[\Xi^-\bar{K}^0]$~~~~~~&$\Gamma^{\text{th}}_{\text{total}}$\\
$|70,^210,1,1,1/2^-\rangle$~~~~~~&7.43~~~~~~&7.82~~~~~~&15.2  \\
$|70,^210,1,1,3/2^-\rangle$~~~~~~&3.51~~~~~~&3.12~~~~~~&6.64\\
\hline\hline
\end{tabular}
\end{table}

\begin{table*}[htbp]
\caption{The predicted partial and total decay widths of the $S$- and $D$-wave states in the $N=2$ shell. $\Gamma^{\text{th}}_{\text{total}}$ stands for the total decay width and $\mathcal{B}$ represents the ratio of the branching fraction $\Gamma[\Xi K]/\Gamma[\Xi(1530) K]$. The unit of widths and masses is MeV.} \label{sdwave-num}
\begin{tabular}{cccccccccccc }\hline\hline
~~~~~~&States~~~~~~&\text{Mass}~\cite{Chao:1980em}~~~~~~&$\Gamma[\Xi K]$ ~~~~~~&$\Gamma[\Xi(1530)K]$ ~~~~~~&$\Gamma[\Omega\eta]$ ~~~~~~&$\Gamma^{\text{th}}_{\text{total}}$ ~~~~~~&$\mathcal{B}$\\
$2S$ \text{wave}~~~~~~&$|70,^210,2,0,1/2^+\rangle$~~~~~~&2190~~~~~~&0.06~~~~~~&2.43~~~~~~&$\cdot\cdot\cdot$~~~~~~&2.49~~~~~~&0.02  \\ ~~~~~~&$|56,^410,2,0,3/2^+\rangle$~~~~~~&2065~~~~~~&1.03~~~~~~&0.96~~~~~~&$\cdot\cdot\cdot$~~~~~~&2.00~~~~~~&1.07  \\
\hline
$1D$ \text{wave}~~~~~~&$|56,^410,2,2,1/2^+\rangle$~~~~~~&2210~~~~~~&51.8~~~~~~&4.53~~~~~~&$\cdot\cdot\cdot$~~~~~~&56.3~~~~~~&11.4  \\ ~~~~~~&$|56,^410,2,2,3/2^+\rangle$~~~~~~&2215~~~~~~&25.8~~~~~~&15.7~~~~~~&$\cdot\cdot\cdot$~~~~~~&41.5~~~~~~&1.64  \\
~~~~~~&$|56,^410,2,2,5/2^+\rangle$~~~~~~&2225~~~~~~&6.58~~~~~~&22.6~~~~~~&0.11~~~~~~&29.2~~~~~~&0.29  \\
~~~~~~&$|56,^410,2,2,7/2^+\rangle$~~~~~~&2210~~~~~~&26.2~~~~~~&1.51~~~~~~&$\cdot\cdot\cdot$~~~~~~&27.7~~~~~~&17.4  \\
~~~~~~&$|70,^210,2,2,3/2^+\rangle$~~~~~~&2265~~~~~~&7.40~~~~~~&11.9~~~~~~&1.60~~~~~~&20.9~~~~~~&0.62  \\
~~~~~~&$|70,^210,2,2,5/2^+\rangle$~~~~~~&2265~~~~~~&0.99~~~~~~&11.6~~~~~~&0.83~~~~~~&13.4~~~~~~&0.08  \\
\hline\hline
\end{tabular}
\end{table*}

Considering the uncertainties of the predicted masses, we plot the variation of the decay properties of the two states as functions of the masses in Fig.~\ref{fig-pwave}. The total decay width of $|70,^210,1,1,1/2^-\rangle$ is about $\Gamma\sim(12-18)$ MeV with the mass varied in the range of (1900-2100) MeV and insensitive to its mass within the considered range. For the state $|70,^210,1,1,3/2^-\rangle$, if it lies below the threshold of $\Xi(1530)K$, its dominant decay mode is $\Xi K$ with a fairly narrow width $\Gamma<8$ MeV. However, if its mass is above the threshold of $\Xi(1530)K$, it mainly decays into $\Xi(1530)K$ channel and its total decay with may reach up to $\Gamma\sim90$ MeV with the mass $M=2100$ MeV.

\subsection{$2S$ wave states in the $N=2$ shell }

In the quark model, there are two $2S$ wave states with $J^P=1/2^+$ and $J^P=3/2^+$, i.e. $|70,^210,2,0,1/2^+\rangle$ and $|56,^410,2,0,3/2^+\rangle$. In Ref.~\cite{Chao:1980em}, their masses were predicted to be about 2.19 GeV and 2.06 GeV~\cite{Chao:1980em}, respectively.
Using these predicted masses, we calculate their partial and total strong decay widths. Our results have been listed in Table~\ref{sdwave-num}. It is found that both $|70,^210,2,0,1/2^+\rangle$ and $|56,^410,2,0,3/2^+\rangle$ are most likely to be the fairly narrow states with a width of $\Gamma\sim2.0$ MeV.

The dominant decay mode of $|70,^210,2,0,1/2^+\rangle$ is $\Xi(1530)K$.
While the dominant decay modes of $|56,^410,2,0,3/2^+\rangle$ are $\Xi K$ and $\Xi(1530)K$, and
the predicted branching fraction ratio is
\begin{eqnarray}
\frac{\mathcal{B}[|56,^410,2,0,3/2^+\rangle\rightarrow \Xi K ]}{\mathcal{B}[|56,^410,2,0,3/2^+\rangle\rightarrow \Xi(1530) K ]}\simeq 1.07.
\end{eqnarray}
From the point of view of the mass and decay width, we can't excluded the first radially excited $\Omega$ state
$|56,^410,2,0,3/2^+\rangle$ as a assignment of the newly observed $\Omega(2012)$ state.

\begin{figure*}[htbp]
\centering \epsfxsize=17.0cm \epsfbox{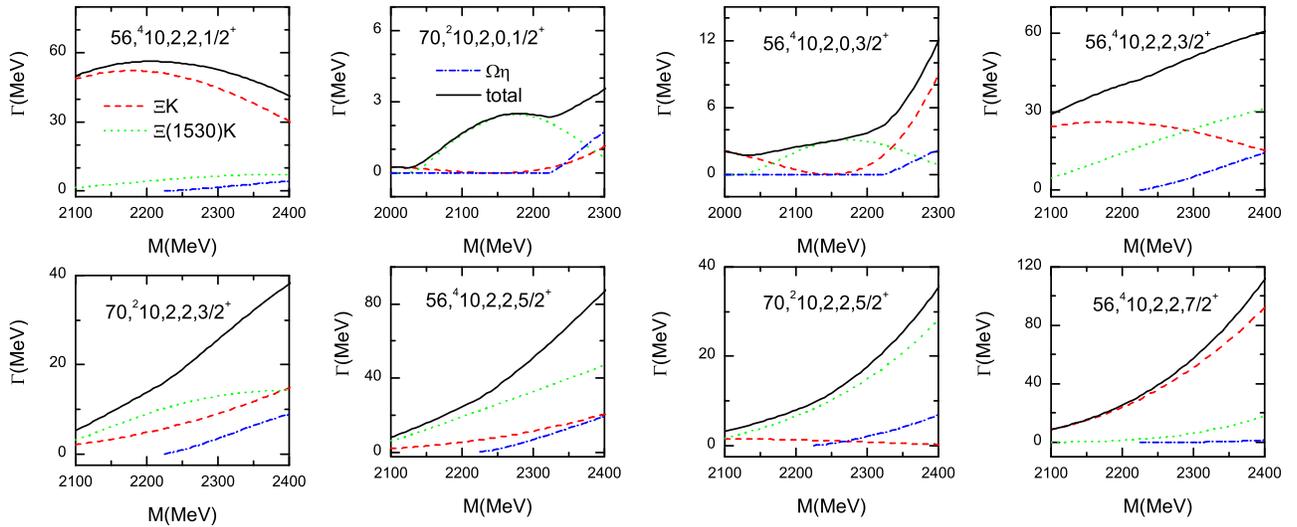} \caption{The
strong decay properties of the $2S$- and $1D$-wave states in the $N=2$ shell. }\label{fig-sdwave}
\end{figure*}

In addition, we also plot the decay widths of $|70,^210,2,0,1/2^+\rangle$ and $|56,^410,2,0,3/2^+\rangle$ as functions of the masses in the range of $M=(2000-2300)$ MeV in Fig.~\ref{fig-sdwave}. The variation curves between the decay widths and the masses of the two states can be obtained from the figure.

\subsection{$1D$ wave states in the $N=2$ shell}

There are six $1D$ wave states according to the quark model classification (see Table~\ref{wfL}). Their masses are estimated to be
in the range of $2.2-2.3$ GeV in various quark models. With the predicted masses from Ref.~\cite{Chao:1980em}, we further analyze the decay properties of the $1D$ wave states in the $N=2$ shell, and collect their strong decay widths in Table~\ref{sdwave-num}. The predicted masses of the $1D$ wave states certainly have a large uncertainty, which may bring uncertainties to our theoretical predictions. To investigate this effect, we plot the total and partial decay widths of these states as functions of the masses in the range of $M=(2100-2400)$ MeV in Fig.~\ref{fig-sdwave} as well.

It is found that the total decay widths of the $1D$ wave states are not broad, they are about $\Gamma\simeq (10-100)$ MeV.
The strong decays of both $|56,^410,2,2,1/2^+\rangle$ and $|56,^410,2,2,7/2^+\rangle$ are governed by the $\Xi K$ mode.
The strong decays of both $|56,^410,2,2,5/2^+\rangle$ and $|70,^210,2,2,3/2^+\rangle$ are governed by the $\Xi(1530) K$ mode.
While the $|56,^410,2,2,3/2^+\rangle$ and $|70,^210,2,2,3/2^+\rangle$ states mainly decay into $\Xi K$ and $\Xi(1530) K$ channels.

It should be mentioned that the $\Omega(2250)$ resonance with a width of $\Gamma=55\pm 18$ MeV listed in PDG~\cite{Patrignani:2016xqp} may be a good
candidate of $|56,^410,2,2,5/2^+\rangle$.
The $\Omega(2250)$ was seen in the $\Xi(1530) K$ and $\Xi^-\pi^+K^-$ channels. The measured mass of
$\Omega(2250)$ is consistent with the quark model predictions~\cite{Capstick:1986bm,Chao:1980em}.
Assigning it as $\Omega(2250)$, the total width is
predicted to be
\begin{eqnarray}
\Gamma_{\mathrm{total}}^{\mathrm{th}}[\Omega(2250)]=36 ~\text{MeV}.
\end{eqnarray}
Its strong decays are dominated by the $\Xi(1530) K$ mode, while the decay rate into
the $\Xi K$ is sizeable. The partial width ratio between
$\Xi(1530) K$ and $\Xi K$ is predicted to be
\begin{eqnarray}
\frac{\mathcal{B}[\Omega(2250)\to \Xi(1530) K]}{\mathcal{B}[\Omega(2250)\to\Xi K]}\simeq 3.2.
\end{eqnarray}
Both the decay width and decay mode are consistent with the observations.

As a whole, the $1D$ wave states are relatively narrow states with a typical width of 10s MeV.
They mainly decay into $\Xi K$ and/or $\Xi(1530) K$ final states. To establish
these missing $1D$ wave states, observations in the both $\Xi K$ and $\Xi(1530) K$
channels are expected to be carried out in future experiments.

\section{Summary}

In the present work, we carry out a systematic study of the OZI allowed two-body strong decays of $\Omega$ resonances up to the $N=2$ shell within the chiral quark model. For the newly observed $\Omega(2012)$ state, we give a possible interpretation in theory. Meanwhile, we give the predictions for the decay properties of the $1D$ wave states, and hope to provide helpful information for searching these missing $\Omega$ states in the future.

The newly observed $\Omega(2012)$ state is most likely to be explained as the $1P$ wave state with $J^P=3/2^-$,
$|70,^210,1,1,3/2^-\rangle$. Meanwhile, with the present information from experiments we can't rule out
the $\Omega(2012)$ as the assignments of the $1P$ wave state $|70,^210,1,1,1/2^-\rangle$ with $J^P=1/2^-$
and the $2S$ wave state $|56,^410,2,0,3/2^+\rangle$ with $J^P=3/2^+$ completely.

The $\Omega(2250)$ resonance listed in PDG may be a good
candidate of the $J^P=5/2^+$ $1D$ wave state $|56,^410,2,2,5/2^+\rangle$.
Generally, the $1D$ wave states are relatively narrow states with a typical width of 10s MeV.
They mainly decay into $\Xi K$ and/or $\Xi(1530) K$ final states. To establish
these missing $1D$ wave states, observations in the both $\Xi K$ and $\Xi(1530) K$
channels are expected to be carried out in future experiments.

\section*{Acknowledgements }
 We would like to thank Shi-Lin Zhu for very
helpful discussions. This work is supported by the National Natural
Science Foundation of China under Grants No. 11775078. This work is also in part supported by China Postdoctoral
Science Foundation under Grant No.~2017M620492.


\end{document}